\documentclass[aps,prx,twocolumn,groupeaddress,superscriptaddress]{revtex4-2}

\usepackage[dvipdfmx]{graphicx}
\usepackage{helvet}
\usepackage{amsmath,amssymb}
\usepackage{bm}
\usepackage{colortbl}
\usepackage{braket}
\begin{document}
\title{Absence of spin susceptibility decrease in a bulk organic superconductor with triangular lattice}

\author{Y. Saitou}
\affiliation{
Department of Applied Physics, Tokyo University of Science, Tokyo 125-8585, Japan}

\author{N. Ichikawa}
\affiliation{
Department of Applied Physics, Tokyo University of Science, Tokyo 125-8585, Japan}

\author{R. Yamamoto}
\altaffiliation{
Present Address: Los Alamos National Laboratory, New Mexico, USA}
\affiliation{
Department of Applied Physics, Tokyo University of Science, Tokyo 125-8585, Japan}

\author{D. Kitamata}
\affiliation{
Department of Applied Physics, Tokyo University of Science, Tokyo 125-8585, Japan}

\author{M. Suzuki}
\affiliation{
Department of Applied Physics, Tokyo University of Science, Tokyo 125-8585, Japan}

\author{Y. Yanagita}
\affiliation{
Department of Applied Physics, Tokyo University of Science, Tokyo 125-8585, Japan}

\author{T. Namaizawa}
\affiliation{
Department of Applied Physics, Tokyo University of Science, Tokyo 125-8585, Japan}

\author{S. Komuro}
\affiliation{
Department of Applied Physics, Tokyo University of Science, Tokyo 125-8585, Japan}

\author{T. Furukawa}
\affiliation{
Department of Applied Physics, Tokyo University of Science, Tokyo 125-8585, Japan}
\affiliation{Institute for Materials Research, Tohoku University, Sendai 980-8577, Japan}

\author{R. Kato}
\affiliation{
Condensed Molecular Materials Laboratory, RIKEN, Saitama 351-0198, Japan}

\author{T. Itou}
\email{tetsuaki.itou@rs.tus.ac.jp}
\affiliation{
Department of Applied Physics, Tokyo University of Science, Tokyo 125-8585, Japan}

\date{\today}

\begin{abstract}

The study of non-$s$-wave unconventional superconductivities in strongly correlated-electron systems has been a central issue in condensed matter physics for more than 30 years. In such unconventional superconductivities, $d$-wave Cooper pairing with antiparallel spins has been often observed in various quasi-two-dimensional (quasi-2D) bulk systems. Interestingly, many theories predicted that the triangular lattice causes the $d$-wave pairing to be unstable and may lead to more exotic pairing such as parallel spin (spin-triplet) pairing.  
Here we focus on a bulk organic triangular-lattice system in which superconductivity emerges near a nonmagnetic Mott insulating phase. We demonstrate, by using low-power nuclear magnetic resonance (NMR) measurements, that the spin susceptibility of the superconducting state retains the normal state value even deep in the superconducting state. This result indicates the possibility that the material exhibits spin-triplet superconductivity. Our finding will bring insights also into understanding the 2D materials with triangular moir\'{e} superlattices that are considered also to show unconventional superconductivities near Mott-like insulating states.
\end{abstract}

\maketitle

\section{Introduction}
Superconductivity in strongly correlated electron systems has been one of the most interesting issues in condensed-matter physics. In general, Cooper pair symmetries in strongly correlated electron systems tend to be non-$s$-waves because electrons are reluctant to occupy the same sites. Among the non-$s$-wave states, $d$-wave superconductivity is most often realized, which has been considered to emerge in various bulk quasi-2D materials including high-$T_{\mathrm{c}}$ cuprates and organic materials~\cite{cuprates,organics}. However, in triangular lattice systems, the six-fold rotational symmetry inherent in the systems causes several $d$-wave states to be degenerate. In this frustrating situation, $d$-wave superconductivity tends to be energetically unstable, which might result in the realization of more exotic superconductivities preserving the rotational symmetry, such as chiral $d+id$-wave ($d_{x^2-y^2}+id_{xy}$-wave) superconductivity and spin-triplet superconductivities~\cite{ES1,ES2,ES3,ES4,ES5,ES6,ES7,ES8,ES10,ES11,ES12,ES13,ES14,ES15,ES16,ES17,ES18,ES19,ES20,ES21,ES22,ES23,ES24,ES25,ES26}, as shown in Fig. 1. These exotic superconductivities can survive even when the lattice is not a perfect equilateral triangle. Indeed, some theoretical works proposed that the chiral superconductivity can be realized even on anisotropic triangular lattices~\cite{Powell1, Powell2}. Despite its importance, correlated-electron superconductors with triangular lattices very seldom occur in nature; indeed, only a few candidates have been theoretically studied, e.g., Na$_{x}$CoO$_{2} \cdot y$H$_{2}$O~\cite{ES1,ES2,ES3,ES4,ES5,ES6,ES7,ES8,ES10,ES11,ES12,ES13,ES14,ES15,ES16,ES17,ES18,ES19,ES20,ES21,ES22,ES23,ES24,ES25,ES26}. However, recent studies reported that 2D magic-angle twisted bilayer/trilayer graphene systems with triangular moir\'{e} superlattices show superconducting states nearby Mott-like insulating states~\cite{Cao1,Cao2}, sparking massive attention. Furthermore, a very recent work has suggested that superconductivity in twisted trilayer graphene is possibly spin-triplet, and the work also stated that further investigations are required to determine the full pairing symmetry in such superconductivities~\cite{Cao3}. Note that, however, possible experimental techniques are very limited for the moir\'{e} graphene superconductivities because of the pure 2D nature; it is difficult to directly observe the spin states of the 2D superconductivities. In contrast, for bulk systems, spin states can be investigated by various methods such as NMR techniques. In this situation, bulk triangular-lattice superconducting materials with strong electronic correlation that can be investigated by extensive experimental techniques are highly desired to reveal the triangular-lattice superconducting physics, which is currently attracting eager interest in moir\'{e} graphene science.

\begin{figure}[ht]
\includegraphics[scale=1.0]{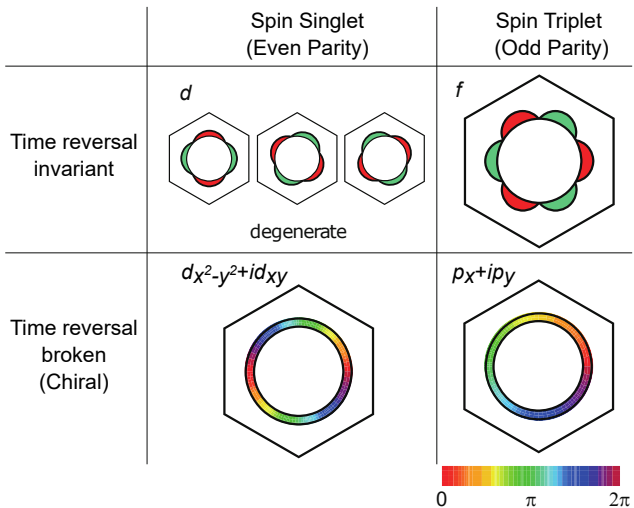}
\caption{Orbital order parameters for possible non-$s$-wave superconductivities on triangular lattice. $d$-wave and chiral $d+id$-wave superconductivities have even-parity and spin-singlet order parameters, whereas $f$-wave and $p+ip$-wave have odd-parity and spin-triplet order parameters.}
\label{Fig. 1}
\end{figure}

In the present study, we focus on a bulk organic triangular-lattice system in which superconductivity is realized nearby a nonmagnetic Mott insulating phase. We demonstrate, by using low-power NMR measurements, that the superconductivity has Pauli paramagnetic spin susceptibility that retains the normal state value even deep in the superconducting state. The studied material is the bulk quasi-2D nearly isotropic triangular lattice system EtMe$_3$P[Pd(dmit)$_2$]$_2$ (space group $P$2$_1$/$m$), where Et = C$_2$H$_5$, Me = CH$_3$, and dmit = 1,3-dithiol-2-thione-4,5-dithiolate, C$_3$S$_5$. The crystal structure is shown in Figs. 2(a) and (b). The family of $X$[Pd(dmit)$_2$]$_2$ ($X$: monovalent closed-shell cation) has a scalene triangular lattice of [Pd(dmit)$_2$]$_2$ dimers, as shown in Fig. 2(b)~\cite{Kato2004}. Although the three transfer integrals [$t_{\mathrm{B}}$, $t_{\mathrm{s}}$, and $t_{\mathrm{r}}$ in Fig. 2(b)] on the three edges of the triangle are in principle different from one another in $X$[Pd(dmit)$_2$]$_2$, they are almost equal in the $X$=EtMe$_3$P system~\cite{Kato2006,Tamura2006}. Therefore, EtMe$_3$P[Pd(dmit)$_2$]$_2$ is considered an electronic system with a nearly equilateral triangular-lattice network. As well as the 2D materials with triangular moir\'{e} superlattices, the present system can also be described by the Hubbard model. Indeed, this system is a Mott insulator at ambient pressure, and below 25 K, shows a spin-gapped nonmagnetic state with dimerization of localized spins [valence-bond-solid (VBS) state]~\cite{Tamura2006,Itou2009}. At a pressure above $\sim$4.5 kbar, a metallic state and superconductivity with $T_{\mathrm{c}} \sim$ 5 K appears through the Mott transition, as shown in Fig. 2(c)~\cite{Kato2006,Itou2009,Shimizu2007,Ishii2007,Tamura2007}. 

Although $d$-waves are believed to be the natural form of superconducting symmetry in strongly correlated electron systems realized in most quasi-2D organic conductors, $d$-wave superconductivity would become unstable in the present system because of the nearly equilateral triangular-lattice nature, which may lead to more exotic superconductivity. To explore this possibility from the viewpoint of microscopic magnetic properties, we performed $^{13}$C-NMR Knight-shift measurements down to 1.5 K for EtMe$_3$P[Pd(dmit)$_2$]$_2$ under pressure, where the Knight shift is an NMR spectral shift due to a hyperfine coupling between nuclear spins and electronic spins and thus directly detects the local electronic spin susceptibility.

\begin{figure*}[ht]
\includegraphics[]{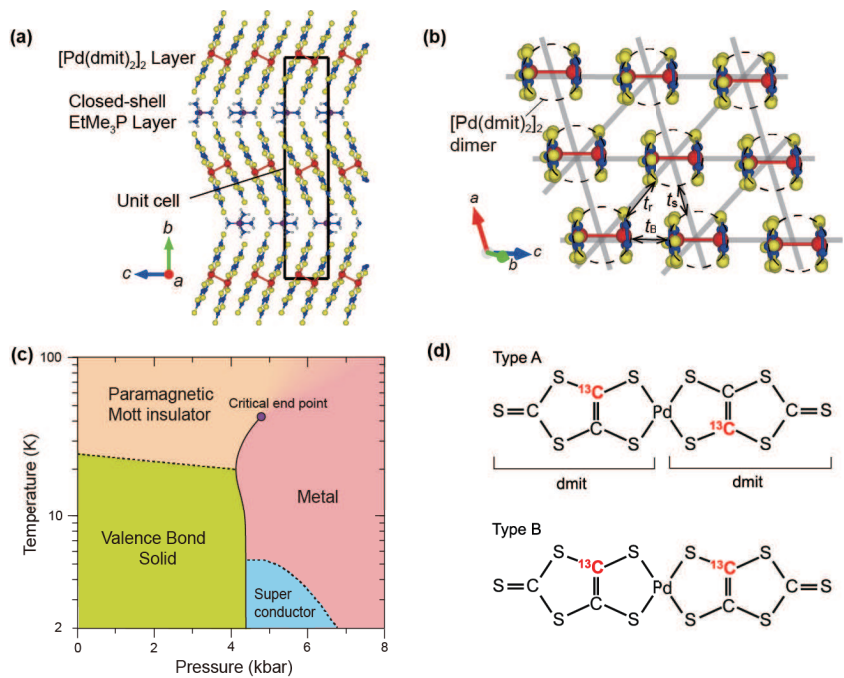}
\caption{(a) Side view of the crystal structure of EtMe$_3$P[Pd(dmit)$_2$]$_2$. Two-dimensional [Pd(dmit)$_2$]$_2$ layers are separated by nonmagnetic layers of the closed-shell monovalent cation EtMe$_3$P. (b) Top view of the crystal structure of the [Pd(dmit)$_2$]$_2$ layer of EtMe$_3$P[Pd(dmit)$_2$]$_2$. The Pd(dmit)$_2$ molecules are strongly dimerized (the pairs are denoted by dashed ovals). One electron with a 1/2 spin is localized on each [Pd(dmit)$_2$]$_2$ dimer in the Mott insulating state at ambient pressure. The arrows ($t_{\mathrm{B}}$, $t_{\mathrm{s}}$, and $t_{\mathrm{r}}$) indicate the transfer integral network between the molecular orbitals of the [Pd(dmit)$_2$]$_2$ dimers. The three transfer integrals are non-equivalent but close to each other in EtMe$_3$P[Pd(dmit)$_2$]$_2$. (c) Pressure-temperature phase diagram of EtMe$_3$P[Pd(dmit)$_2$]$_2$. (The values of the horizontal axis indicate pressures applied at room temperature and actual pressures at low temperatures decrease by 1.5--2 kbar from the room-temperature values.~\cite{Murata1997}) (d) $^{13}$C-enriched Pd(dmit)$_2$ molecules for the present NMR measurements. Only one side of each inner double bond is enriched, as shown in the upper ($^{13}$C-enriched type A) and lower (type B) molecules, which are equally contained in the present samples.}
\label{Fig. 2}
\end{figure*}

Recently, it was reported that radio-frequency (RF) pulses in NMR measurements cause heating problem in Sr$_2$RuO$_4$~\cite{Pustogow2019, Ishida2020}, which had been considered a promising candidate for a triplet superconductor~\cite{Ishida1998}, and that special care to avoid the heating problem is necessary to obtain reliable data on Knight shifts in the superconducting state. Thus, to avoid the heating problem, we performed $^{13}$C-NMR measurements using low-power RF pulses of less than 1 $\mu$J, while typical energies for usual NMR measurements are $\sim$100 $\mu$J. The measurements were done for four single crystals (sample \#1-4, see Appendix). We also investigated the pulse energy dependence of the $^{13}$C-NMR spectrum and the Knight shift for sample \#4 at 1.7 K, which is well below the superconducting transition temperature ($T_{\mathrm{c}}$ = 4.3 K under the magnetic field for the NMR measurements). By analyzing these results, we succeeded in obtaining reliable NMR data completely free from the heating problem and found that the Knight shift does not decrease in the superconducting state. This means that the spin susceptibility retains the normal state value in the superconducting state of EtMe$_3$P[Pd(dmit)$_2$]$_2$, indicating the possibility of spin-triplet superconductivity.

\section{Results}
\subsection{Basic properties of superconductivity}

To know the basic properties of the superconductivity of the present material, we measured the ac susceptibility for the four samples under a pressure of $\sim$5.0 kbar, under which superconductivity is realized, by analyzing the resonance frequency of the NMR LC tank circuit (see Appendix).

Figure 3 shows the temperature dependence of the ac susceptibility $\chi_\mathrm{ac}$ for sample \#1-4 at $H$ = 0 and 1.70--2.25 T. The magnetic fields of 1.70--2.25 T were applied exactly parallel to the conducting $ac$ plane with an accuracy of  $\pm0.5^\circ$. As shown in Fig. 3, the diamagnetic susceptibility owing to the Meissner effect is observed as an increase in $- \chi_\mathrm{ac}$. The value of $T_{\mathrm{c}}$ at $H = 0$ is approximately 4.5 K, and the four samples show slightly different $T_{\mathrm{c}}$, probably owing to slight pressure differences. We also estimated the in-plane and out-of-plane coherence lengths ($\xi_\parallel(0) \sim 120$ $ \mathrm{\AA}$ and $\xi_\perp(0) \sim 22$ $\mathrm{\AA}$) and magnetic penetration lengths ($\lambda_\parallel (0) \lambda_\perp (0) \sim 0.8$ $\mu$m$^2$) from the ac susceptibility data (see Supplemental Material~\cite{Supplement}). Note that the diamagnetic signal increases rather gradually, as shown in Fig. 3. This is basically because of the magnetic penetration effect. Around $T_{\mathrm{c}}$, the penetration length tends to diverge and becomes comparable to or longer than the sample thickness of $\sim$50 $\mu$m, which suppresses the diamagnetic signal.

\begin{figure*}[ht]
\includegraphics[width=\hsize]{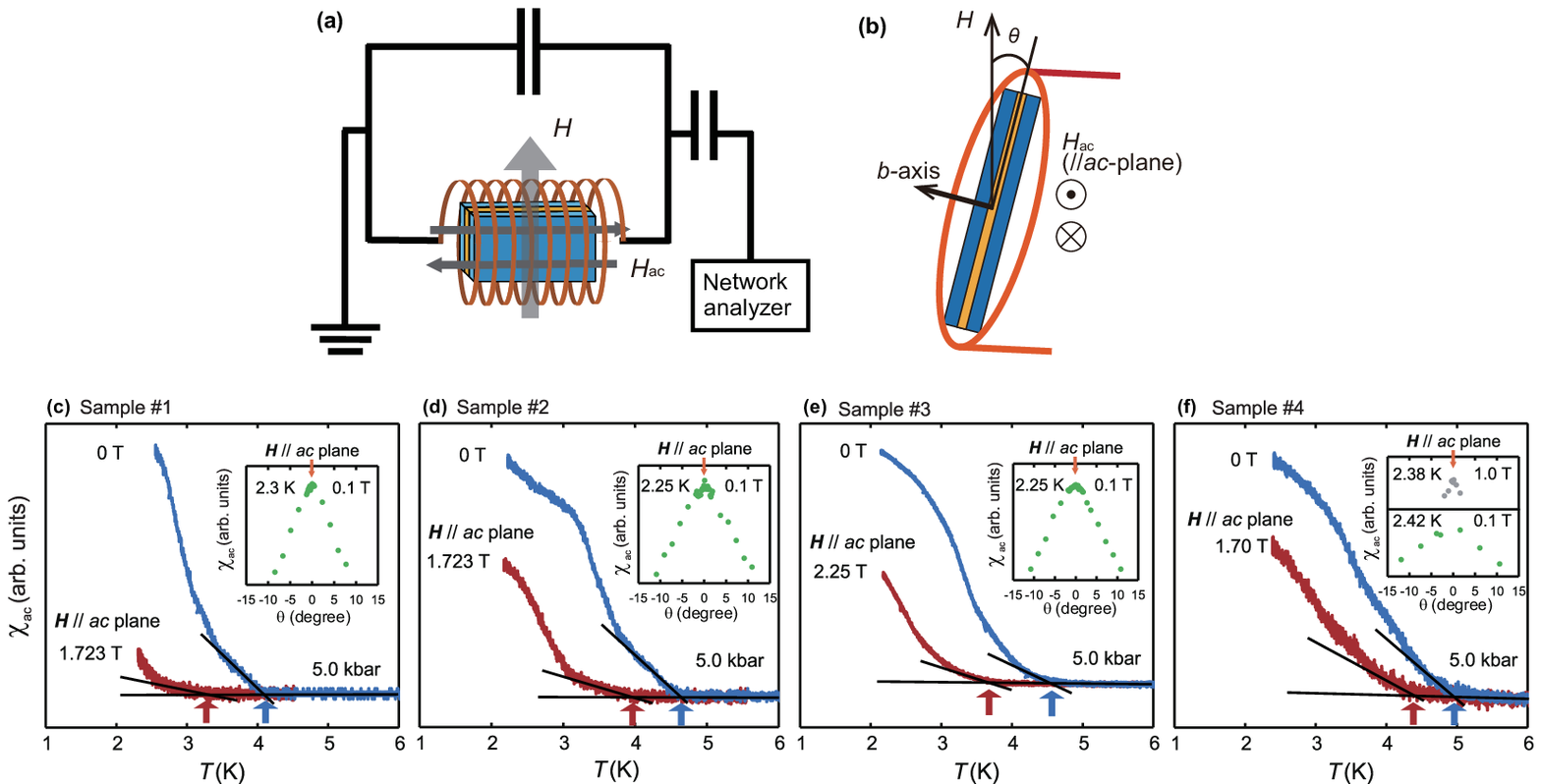}
\caption{(a) LC tank circuit to measure ac susceptibility. Static
magnetic field $H$ was applied perpendicular to the ac field $H_\mathrm{ac}$.
(b) Schematic for the configuration of $H$ and $H_\mathrm{ac}$. The angle between
the directions of the two-dimensional layers ($ac$ plane) and $H$ is defined as $\theta$,
which was varied by a rotation mechanism. (c)-(f) Temperature dependence of the ac susceptibility $\chi_\mathrm{ac}$ of EtMe$_3$P[Pd(dmit)$_2$]$_2$ under a pressure of $\sim$5.0 kbar at $H = 0$ (blue data) and 1.70--2.25 T (red data). The magnetic fields of 1.70--2.25 T were applied exactly parallel to the conducting $ac$ plane ($\theta = 0^\circ$) and the same as the fields under which NMR measurements shown in Figs.~\ref{Fig. 4} and~\ref{Fig. 5} were done. The arrows indicate the onset temperature of the increase in the diamagnetic signal of
the ac susceptibility. The insets show $\theta$ dependence of the ac susceptibility at $H = 0.1$ T (green dots) and 1.0 T (gray dots).}
\label{Fig. 3}
\end{figure*}

\subsection{$^{13}$C-NMR of superconducting phase} 

\begin{figure*}[ht]
\includegraphics[]{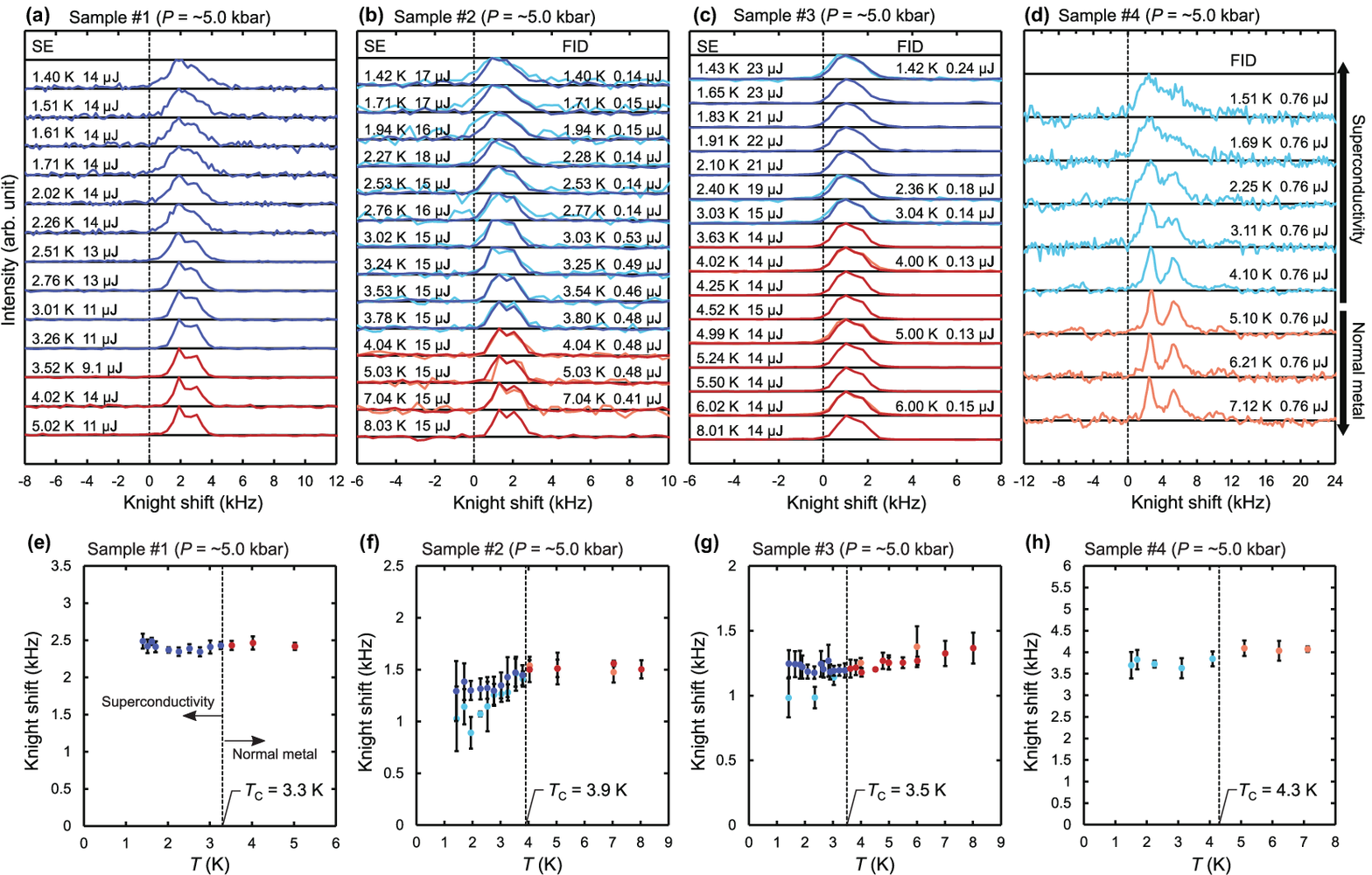}
\caption{(a)-(d) Temperature dependence of the $^{13}$C-NMR spectra for sample \#1-4 under the pressure of $\sim$5.0 kbar. The darker-colored lines in (a)-(d) show data obtained using the SE method, and the lighter-colored lines in (b)-(d) show data obtained using the reduced-power FID method. (e)-(h) Temperature dependence of the $^{13}$C-NMR Knight shifts obtained from the data shown in (a)-(d). The darker-colored symbols in (e)-(h) show data obtained using the SE method, and the lighter-color symbols in (f)-(h) show data obtained using the reduced-power FID method. }
\label{Fig. 4}
\end{figure*}

Here we show the main result of the present work, the $^{13}$C-NMR data under the pressure of $\sim$5.0 kbar, under which superconductivity emerges.
The NMR measurements were performed under exactly the same pressure and magnetic field conditions as the ac susceptibility measurements under applied field (the red data in Fig. 3); hence, the external magnetic fields $H$ (sample \#1: 1.723 T, \#2: 1.723 T, \#3: 2.25 T, \#4: 1.70 T) were applied exactly parallel to the conducting $ac$ plane. 
Figure~\ref{Fig. 4} shows the temperature dependence of the $^{13}$C-NMR spectra and the Knight shift (defined by the spectral centers of gravity) for sample \#1-4 under the pressure down to 1.5 K. The superconducting transition temperature $T_{\mathrm{c}}$ is determined from the onset temperature of the increase in the diamagnetic signal observed in the ac susceptibility (shown in Fig. 3). Note that we also observed a decrease in $(T_{1}T)^{-1}$ below $T_{\mathrm{c}}$ (see Supplemental Material~\cite{Supplement}), where $T_{1}$ is the spin-lattice relaxation time. At a glance, the Knight shift behavior looks unusual because the Knight shifts do not decrease in the superconducting phase unlike the case of singlet superconductivity. Before looking at the details of this peculiar behavior, we have to check the two important technical points: (1) the origin of the Knight shift and (2) the heating problem mentioned before.

\begin{figure*}[ht]
\includegraphics[]{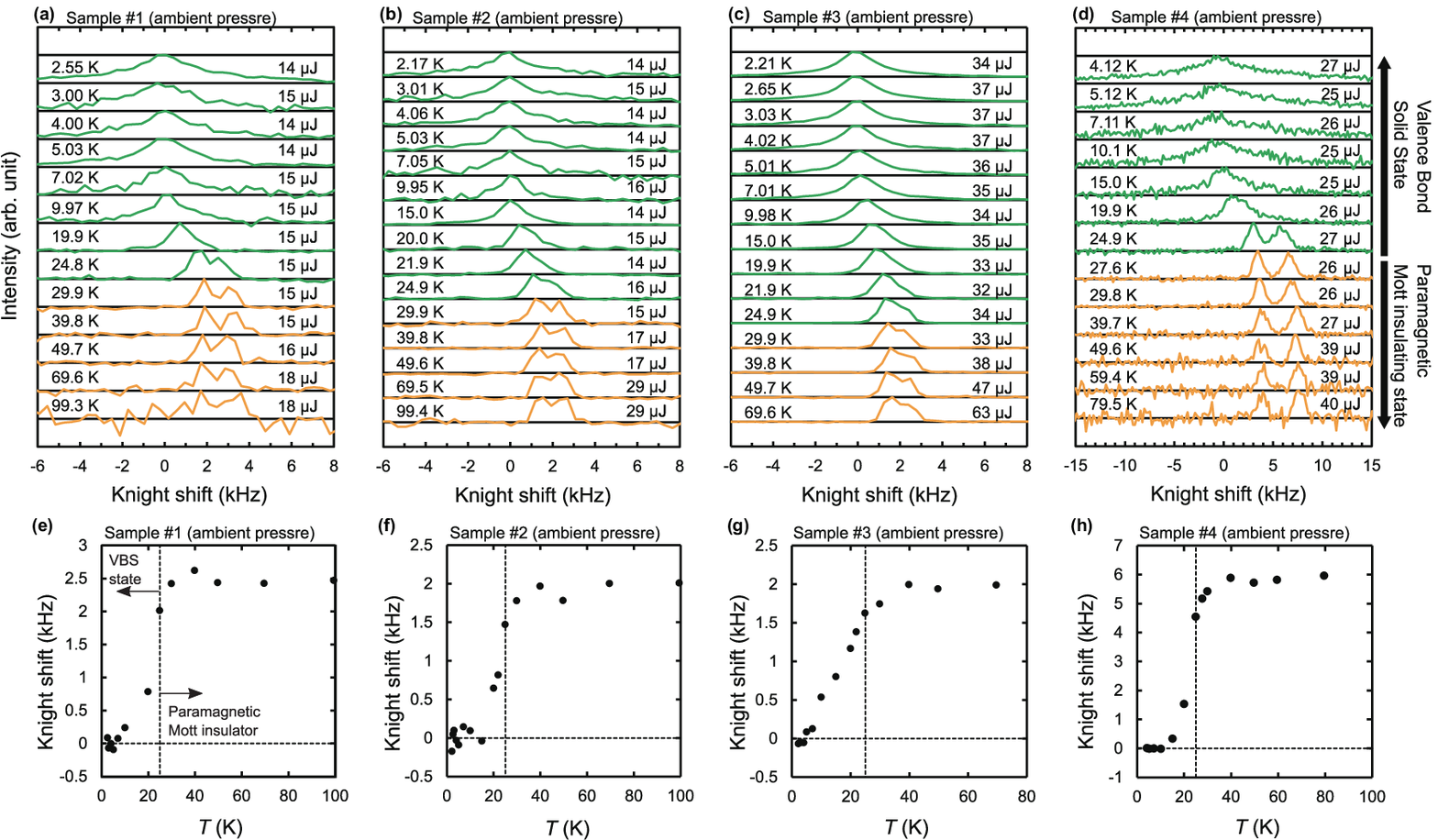}
\caption{(a)-(d) Temperature dependence of the $^{13}$C-NMR spectra for sample \#1-4 at ambient pressure. (e)-(h) Temperature dependence of the $^{13}$C-NMR Knight shifts (spectral centers of gravity) obtained from the data shown in (a)-(d). The horizontal dashed lines indicate the Knight shift origins obtained from the spectral centers of gravity at low temperatures. The vertical dashed lines indicate the valence-bond-solid transition temperature. All measurements were performed using the spin echo (SE) method, with a total pulse energy of several tens of $\mu$J. 
}
\label{Fig. 5}
\end{figure*}

To evaluate a Knight shift quantitatively, unambiguous determination of the Knight shift origin is crucial but often difficult experimentally.
Fortunately, the Knight shift origins shown in Fig.~\ref{Fig. 4} can be clearly determined from the $^{13}$C-NMR data at ambient pressure, under which the nonmagnetic VBS insulating ground state is realized at low temperatures~\cite{Tamura2006,Itou2009}.
As shown in Figs.~\ref{Fig. 5}(a-d), in the ambient-pressure Mott-insulating phase, the spectral centers of gravity of all samples shift to lower frequencies below 25 K. This is because the spin susceptibility below the VBS transition temperature vanishes owing to the opening of the spin gap. In the low-temperature limit, where the spin susceptibility completely vanishes, the Knight shifts become exactly zero; hence, the residual shifts are caused only by chemical shifts. Note that the direction of a magnetic field and thus the value of the chemical shift for each sample are the same for both measurements under ambient pressure and $\sim$5.0 kbar. Thus, the Knight shift origins at $\sim$5.0 kbar in Fig.~\ref{Fig. 4} can be defined as the low-temperature limit of the spectral centers of gravity at ambient pressure (Figs.~\ref{Fig. 5}(e-h)). Note that the Knight shift values of the four samples at high temperatures are different. This is because the Knight shift varies depending on the in-plane angle of an external magnetic field even when the magnetic field is exactly parallel to the conducting $ac$ plane. As seen in Figs.~\ref{Fig. 4} and~\ref{Fig. 5}, the Knight shift value is the largest for \#4 (see Appendix and Supplemental Material~\cite{Supplement} for details). 

As explained above, the heating problem caused by RF pulses must be handled. Thus, the present measurements under pressure for sample \#2-4 were performed not only using the conventional spin echo (SE) method but also using the reduced-power free induction decay (reduced-power FID) method. The reduced-power FID method can decrease the total RF-pulse energy much more than the conventional SE method by lowering the applied voltage of the NMR pulse (see Appendix). While the present total RF-pulse energy in the SE method was 10-25 $\mu$J, we succeeded in decreasing the total RF-pulse energy to less than 1 $\mu$J in the reduced-power FID method. In Fig.~\ref{Fig. 4}, the darker-colored lines and symbols show data obtained by the SE method, and the lighter-color shows data obtained by the reduced-power FID method. Importantly, as discussed later, the result of the pulse energy dependence of the NMR spectra guarantees that the heating effect is completely negligible in the reduced-power FID measurements for sample \#4, which has the largest Knight shift among the four measured samples. Therefore, the data of the reduced-power FID measurements for \#4 are the most reliable for discussing the Knight shift.

Now we have solved the two technical problems; we move on to the details of the Knight shift behavior. As shown in Fig.~\ref{Fig. 4}, the Knight shift does not decrease below $T_{\mathrm{c}}$ for any of the samples. This indicates that the spin susceptibility of the superconducting state is temperature independent and retains the normal state value.  
This behavior is in contrast to the case of an ideal singlet superconductor, in which the Knight shift should rapidly decrease toward the Knight-shift origin that is shown by the dashed lines in Figs.~\ref{Fig. 4}(a-d). Looking more closely, seeming slight decreases in the Knight shift appear to be observed for sample \#2 and \#3, although the slight decreases are much less than those expected for an ideal singlet superconductor. This possible small shift is well explained quantitatively by the diamagnetic shift, which does not indicate a decrease in the spin susceptibility. (The details are provided in Supplemental Material~\cite{Supplement}.) Note that  decrease is almost undetectable in the Knight shift for sample \#4, which shows the largest Knight shift and thus is most sensitive to the spin susceptibility 

To confirm that the NMR spectra below $T_{\mathrm{c}}$ indeed come from the superconducting state, here we focus on the spectral broadening below $T_{\mathrm{c}}$. As a temperature is decreased below $T_{\mathrm{c}}$, the NMR spectra show slight broadening not only on the lower-frequency side but also on the higher-frequency side. This broadening is explained by the local field distribution (the Redfield pattern~\cite{Redfield}) associated with a vortex lattice in the superconducting state. This result provides another evidence that the samples are indeed in the superconducting state, together with the decreases in the ac magnetic susceptibility (Fig.~3) and $(T_{1}T)^{-1}$ (see Supplemental Material~\cite{Supplement}). Here, we note that the superconducting state has no significant inhomogeneity. This is because the NMR spectra show typical broadening characteristic of bulk superconductivity and the spin-lattice relaxation curves in the superconducting state are still almost single exponential (see Supplemental Material Fig.~S8~\cite{Supplement}). The single-exponential relaxation also indicates that the vortex core contribution to the NMR signal is sufficiently small. 

Figure 6 shows the pulse energy dependence of the $^{13}$C-NMR spectrum and Knight shift for sample \#4 under a pressure of $\sim$5.0 kbar measured using the reduced-power FID method. The measurements were performed at 1.7 K, which is sufficiently lower than the superconducting transition temperature under the magnetic field for the present NMR measurements ($T_{\mathrm{c}} = 4.3$ K and $T/T_{\mathrm{c}} = 0.4$). The spectrum measured at a relatively high pulse energy of 9.7 $\mu$J is slightly narrower than the other spectra. This indicates that the 9.7 $\mu$J pulse energy partially ruins the local field distribution effect caused by the vortex state. Thus, the heating effect more or less exists and the superconducting state is partially broken in the 9.7 $\mu$J measurement for sample \#4. On the contrary, the spectrum obtained at 0.76 $\mu$J is not different from that at the lowest pulse energy of 0.27 $\mu$J within experimental uncertainty. Considering this feature and that the heating effect should be proportional to the pulse power, we can safely conclude that the NMR data obtained at a pulse energy of 0.76 $\mu$J or less for sample \#4 are free from the heating problem and undoubtedly reflect the properties of the superconducting state.

\begin{figure}[hb]
\includegraphics[]{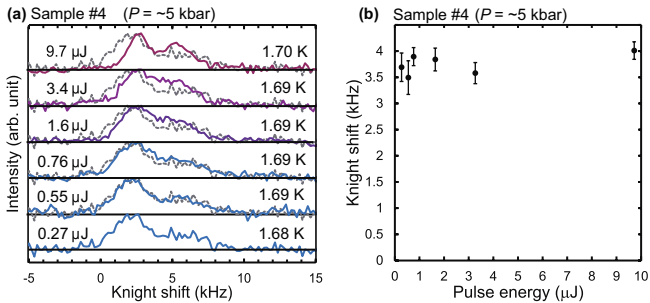}
\caption{(a) Pulse energy dependence of the $^{13}$C-NMR spectrum for sample \#4 measured by the reduced-power FID method at $\sim$1.7 K, $\sim$5.0 kbar. The spectrum of the lowest pulse energy of 0.27 $\mu$J is indicated in each row for comparison (gray dashed lines). (b) Pulse energy dependence of the $^{13}$C-NMR Knight shift obtained from the data shown in (a).}
\label{Fig. 6}
\end{figure}

\section{Discussion}
 
From the above temperature and pulse energy dependence of the $^{13}$C-NMR data, we succeeded in obtaining reliable NMR data for the superconducting state without the heating problem. The data show no obvious decrease in the Knight shift below $T_{\mathrm{c}}$, indicating that the spin susceptibility of the superconducting state retains the normal state value in the temperature range down to 1.5 K ($T/T_{\mathrm{c}}$ $\sim$ 0.35). These results suggest that this superconductivity on a triangular lattice is not simple spin-singlet but unconventional.

Even in the case of spin-singlet superconductivities, it was proposed that the spin-flip scattering by defect or disorder could weaken the Knight shift decrease expected in spin-singlet pairing when the spin-orbit interaction is strong enough to yield the spin scattering mean free path comparable to the coherence length~\cite{Anderson,Miyake}. However, this scenario is clearly unlikely in the present system because the spin-orbit interaction is too weak. The wave function of the conduction electrons is composed of the highest occupied molecular orbital of Pd(dmit)$_2$, whose density lies only on the dmit ligands [Fig. 2(d)] with a negligible contribution from Pd orbitals~\cite{Kato2004}. Because the wave function spreads only on light elements, the spin-orbit interaction in the present system is sufficiently weak, and thus the spin scattering mean free path is expected to be orders of magnitude longer than the total mean free path ($l_\parallel \sim 100-1000$ $\mathrm{\AA}$, Supplemental Material~\cite{Supplement}) and thus the in-plane coherence length ($\xi_\parallel(0) \sim 120$ $\mathrm{\AA}$). 
Therefore, the $^{13}$C-NMR results suggest the possibility that EtMe$_3$P[Pd(dmit)$_2$]$_2$ is a spin-triplet superconductor under pressure. If this is the case, then this organic system has considerably high $T_{\mathrm{c}}$ among the spin-triplet superconductor candidates~\cite{Ishida, Kohori, Tou2, Ott, Saxena, Hardy, Hattori, Huy, Nakamine, Ran, Matano, Lee, Kitagawa,Yang}.

Note that the locking of the $\bm{d}$-vector under magnetic field does not work in EtMe$_3$P[Pd(dmit)$_2$]$_2$ when triplet superconductivity is realized. This is because, as explained above, the spin-orbit interaction is too weak in this system.

Possible representative triplet superconductivities that can be realized in triangular-lattice correlated electron systems are $f$-wave and chiral $p+ip$-wave states, which preserve the six-fold rotational symmetry in the triangular lattice. Because EtMe$_3$P[Pd(dmit)$_2$]$_2$ is a correlated electron system with a nearly regular triangular transfer network, such an exotic triplet superconductivity may be realized. Note that the orbital part of the chiral $p+ip$-wave state is time-reversal broken, while that of the $f$-wave 
($f_{x^3-3xy^2}$ or $f_{y^3-3x^2y}$) state is time-reversal invariant. Further studies to examine whether spontaneous time-reversal symmetry breaking is realized, and to investigate the behavior of the spin susceptibility at ultra-low temperatures, will give clues toward complete understanding of the present superconductivity. Our finding will also bring insights into the unified physics of the exotic superconductivities in bulk quasi-2D triangular-lattice systems and the 2D moir\'{e} triangular-superlattice materials under intense debate.

\section{Conclusion}

We measured the $^{13}$C-NMR Knight shift without the heating effect in the superconducting state realized in EtMe$_3$P[Pd(dmit)$_2$]$_2$ under pressure, which is a correlated electron system with a nearly regular triangular transfer network. The Knight shift shows no decrease in the superconducting state down to 1.5 K. This indicates that the spin susceptibility does not disappear in the superconducting state, suggesting the possibility of spin-triplet superconductivity.

\begin{acknowledgments}
This work was supported in part by JSPS KAKENHI (Grant Nos. 25287082, 25220709, 16H06346, and 18H05225).
\end{acknowledgments}

Y.S. and N.I. contributed equally to this work.

\appendix*
\section{Methods}
\textbf{Sample preparation.}
We prepared single crystals of $^{13}$C-enriched EtMe$_3$P[Pd(dmit)$_2$]$_2$ using an aerial oxidation method. In the crystals, the inner carbons of the Pd(dmit)$_2$ molecule are enriched with $^{13}$C isotopes. Only one side of each inner double bond is enriched, as shown in the upper and lower molecules in Fig. 2(d), which are contained equally in the present samples. In this work, we performed experiments for four different single crystals (\#1-4). All the crystals are platelike with a typical area of $\sim$1 mm$^2$ in the conducting $ac$ plane and a typical thickness of $\sim$50 $\mu$m along the $b$ axis. (Note that the definition of the crystal axes in this paper is based on the room temperature structure.)

\vspace{\baselineskip}

\textbf{Applying pressure.}
We inserted a single crystal of EtMe$_3$P[Pd(dmit)$_2$]$_2$ into a coil, which typically has dimensions of $\sim 1 \times 0.5 \times 2$ mm, and packed the arrangement into a Teflon capsule filled with a pressure medium (Daphne 7373 oil). We applied a pressure of $\sim$5.0 kbar at room temperature with a BeCu clamp cell. The pressure was estimated based on the external force applied at room temperature. Note that at low temperatures, the pressure decreases by 1.5--2 kbar from that at room temperature~\cite{Murata1997}. 

\vspace{\baselineskip}

\textbf{Measurements of ac susceptibility.}
Before performing the main NMR measurements, we measured the ac susceptibility for the four samples under a pressure of $\sim$5.0 kbar, under which superconductivity is realized, by analyzing the resonance frequency $f$ of the NMR LC tank circuit [Fig. 3(a)]; that is, we used the same probe for the ac susceptibility measurements and the NMR measurements. The experimental conditions were the same as those in our previous work, and for a thorough description of the experimental details, see Ref.~\cite{Yamamoto2018}. A brief summary is provided below. 

We measured $f$ using a network analyzer (Agilent Technologies E5061A). The ac field $H_\mathrm{ac}$, which was produced by the ac electric current generated by the network analyzer and flowing through the coil, was applied nearly parallel to the conducting $ac$ layers. In addition to $H_\mathrm{ac}$, an external static magnetic field $H$ was applied perpendicular to $H_{ac}$ with a superconducting magnet. Because $f$ is proportional to the inverse of the square root of the coil inductance, the relation between $f$ and the ac susceptibility $\chi_\mathrm{ac}$ (arbitrary units) is denoted by
\begin{equation}
 -\chi_\mathrm{ac} \propto \left(1-\frac{f_0^2}{f^2}\right),
\end{equation}
where $f_0$ is the resonance frequency in the normal state~\cite{Yamamoto2018}. The angle between the directions of the 2D layers ($ac$ plane) of EtMe$_3$P[Pd(dmit)$_2$]$_2$ and $H$ is defined as $\theta$ [Fig. 3(b)], and $\theta = 0^\circ$ means that $H$ is exactly parallel to the 2D layers. The pressure cell was rotated by an angle $\theta$ in the range within $\theta < \pm 15^\circ$ with a rotation pitch of $0.18^\circ$.

The insets of Figs. 3(c)-(f) show the angle $\theta$ dependence of the ac susceptibility at $H = 0.1$ T and 1.0 T. Because EtMe$_3$P[Pd(dmit)$_2$]$_2$ is an anisotropic three-dimensional superconductor, $- \chi_\mathrm{ac}$ shows a simple peak structure without depression due to the lock-in effect even when $H$ is exactly parallel to the $ac$ plane~\cite{Yamamoto2018}. Therefore, we determined the angle at which $- \chi_\mathrm{ac}$ is maximized at $\theta = 0^\circ$ (corresponding to the angle where $H$ is exactly parallel to the $ac$ plane). The red data under the magnetic fields in the main panels in Figs. 3(c)-(f) were obtained in this angle setting.

\vspace{\baselineskip}

\textbf{Measurements of $^{13}$C-NMR.}
After the ac susceptibility measurements, we performed $^{13}$C-NMR measurements under the pressure of $\sim$5.0 kbar (exactly the same as the pressure for the ac susceptibility measurements) and ambient pressure. First, using the same experimental setup for the above-mentioned ac susceptibility experiments, we performed $^{13}$C-NMR measurements under the same pressure and magnetic field as the red data in Fig. 3. The external magnetic field $H$ (\#1: 1.723 T, \#2: 1.723 T, \#3: 2.25 T, \#4: 1.70 T) was set in the direction of $\theta = 0^\circ$. After completion of the $^{13}$C-NMR measurements under the pressure of $\sim$5.0 kbar, we carefully released the pressure in the clamp cell. Then, we attached the clamp cell to the probe again in exactly the same way as the measurements under pressure, and performed $^{13}$C-NMR measurements again at ambient pressure. Thus, the direction of $H$ is the same for both measurements under ambient pressure and the $\sim$5.0 kbar pressure, and is exactly parallel to the 2D layers. We note that, as described below, the in-plane direction of $H$ differs between the measurements for sample \#1-4.

Because EtMe$_3$P[Pd(dmit)$_2$]$_2$ is an organic material with negligible spin-orbit coupling and thus the spin susceptibility is isotropic, the internal magnetic field caused by spin magnetization can be written as $\Delta \bm{H}_{\mathrm{spin}} = \mathbb{A} \chi \bm{H}$, where $\mathbb{A}$ is the hyperfine coupling tensor, $\chi$ is the scalar spin susceptibility, and $\bm{H}$ is the external static magnetic field. The Knight shift $K$ reflects the component parallel to $\bm{H}$ of the internal magnetic field, and thus $K$ is denoted by

\begin{equation}
 K = \gamma \left( \Delta \bm{H}_{\mathrm{spin}} \cdot \frac{\bm{H}}{H} \right) = \gamma \chi \: (\mathbb{A} \bm{H}) \cdot \frac{\bm{H}}{H} \: .
\end{equation}
This indicates that the anisotropy of the Knight shift is caused only by that of the hyperfine coupling tensor. Thus, although the spin susceptibility is isotropic, the Knight shift value changes depending on the in-plane direction of the field. In our experiments, the field was applied in a direction approximately perpendicular to the $a + c$ axis for sample \#1-3, whereas it was applied approximately parallel to the $c$ axis for sample \#4. Consequently, the Knight shift value of \#4 is several times larger than that of \#1-3 (see Supplemental Material~\cite{Supplement}). Thus, the data for \#4 is the most reliable.

The $^{13}$C-NMR data were basically obtained using the conventional SE method. In this SE method, the NMR spectra were obtained by a Fourier transformation of the spin-echo signals following a double pulse sequence, in which the first and second pulses were set to 90$^\circ$ and 180$^\circ$ widths, respectively. In the experiments in the superconducting state, we lowered the pulse power as much as possible, and the total energy of the two pulses were typically 10--25 $\mu$J, which is almost an order of magnitude smaller than the typical energies ($\sim100$ $\mu$J) for the usual NMR measurements. Moreover, to decrease the heating effect thoroughly, we also performed the reduced-power FID method when measuring the superconducting state for sample \#2-4. In this reduced-power FID method, the NMR spectra were obtained by a Fourier transformation of the FID signals following a single pulse. We set the pulse to a $\sim$10$^\circ$ or less width to reduce the pulse power thoroughly, whereas the pulse is generally set to a 90$^\circ$ width in usual FID measurements. Although this reduced-power FID method causes NMR signals to be very weak, we eventually succeeded in obtaining reliable spectra with sufficient signal-to-noise ratio after quite a long experimental time, as shown in Figs.~\ref{Fig. 4} and~\ref{Fig. 6}. In this reduced-power FID method, we decreased the NMR pulse power down to 0.13--0.27 $\mu$J, which is almost three orders of magnitude smaller than the typical energies for usual NMR measurements.

\end{document}